\documentstyle{article}
\begin{document}
{\large
\begin{center}
{\Large \bf GENERATION OF THE ELECTROSTATIC FIELD \\
IN THE PULSAR MAGNETOSPHERE PLASMA}
\vskip 23pt
{\large  T.A. Kahniashvili, G.Z.~Machabeli and I.~S.~Nanobashvili}
\vskip 7pt
{\it Department of Theoretical Astrophysics, Abastumani Astrophysical\\
Observatory, A.Kazbegi ave.$2^a$, 380060 Tbilisi, Republic of Georgia\\}
\end{center}
\vskip 7pt
\baselineskip 18pt
The behaviour of a relativistic electron-positron plasma in the pulsar
magnetosphere is investigated. The equation of the motion of
the magnetospheric plasma is discussed, from which it follows that, if the
plasma particle radial velocity ${V_r}>{c/{\sqrt 2}}$ (where c is the
speed of light), the centrifugal acceleration changes its sign and the
particle braking begins. The stability of the magnetospheric plasma
with respect to the radially oriented potential perturbations is
discussed and the possibility of the electrostatic field generation in
this plasma along the pulsar magnetic field lines is shown. \\
PACS numbers: 97.60.Gb, 94.30.-d, 94.30.Gm, 94.30.Kq
\newpage
\begin{center}
{\large \bf {I. INTRODUCTION}}
\end{center}

The~~investigation~~of~~the~~pulsar~~magnetosphere is of great interest
(see, for \\ example [1-9]). For the determination of the magnetosphere
structure it is very important to study the physical processes in plasma of
magnetosphere.
After the pioneer papers [1,2] it is assumed
that, because of the pulsar corotation with its magnetic field, the
electric field is generated, which has the nonzero component along
the magnetic field. The electric field ejects particles from the
pulsar surface and accelerates them up to relativistic velocities.
The particles moving along curved magnetic field lines radiate
${\gamma}$-quanta and, when their energy ${{\varepsilon}_{\gamma}}$
exceeds electron's doubled rest energy ${{2}{m}{c^2}}$
${({{{\varepsilon}_{\gamma}}{>}{{2}{m}{c^2}}})}$,
${\gamma}$-quantum decays into an
electron-positron pair. This pair is also accelerated in the electric field
and ${\gamma}$-quanta appear again, which again
decay into an electron-positron pairs, etc. This cascade process causes a fast
filling of the magnetosphere with the relativistic electron-positron plasma,
which, in its turn, screens the electric field generated by the pulsar
rotation. In the previous papers (see, for example [4,5])
the strongly turbulized electron-positron plasmas were investigated, but
the process of
the turbulization was not discussed. We investigate the possible ways of the plasma
turbulization, so we discuss the processes in the approximation of the weak
turbulence.

In the first section of the present paper the equation of the motion for the
magnetospheric plasma in the zeroth approximaton of the weak turbulence is
discussed, from which it follows that, if the plasma particle radial
velocity ${V_r}>{c/{\sqrt 2}}$ (where c is the speed of light), the
centrifugal acceleration changes its sign and the particle bra\-king begins.

In the second section the equations for the perturbed quantities (in
the first approximation of the weak turbulence) are presented. Also the
stability of the magnetospheric plasma with respect to the radially
oriented potential perturbations is discussed and the possibility of
the electrostatic field generation in this plasma along the pulsar
magnetic field lines is shown. \\
\newpage
\begin{center}
{\large \bf {II.THE EQUATION OF THE MOTION OF THE MAGNETOSPHERIC PLASMA}}
\end{center}

The equation of the motion of the magnetospheric plasma particles was
discussed in paper [10].
We use the perpendicular rotator model of the pulsar magnetosphere and treat
only the polar cap. The magnetic field lines
are considered as the radial straight lines located in the plane, which is
perpendicular to the pulsar rotation axis. This assumption is
justified, because we discuss the processes in the magnetospheric
layer, the thickness of which is much less than the curvature radius
of the magnetic
field lines. The magnetospheric plasma particles move along the pulsar
magnetic field lines and also corotate  with them, because the
field lines are frozen in the plasma.
The electric field, generated by the pulsar
rotation together with its magnetic field, is screened by the magnetospheric
plasma.

It is convenient to begin the discussion of the plasma particle motion
in the noninertial frame of a rotating magnetic field line, which is
described by the metric:

$$
{{{d}{S^2}}{=}{{-}{(1-{{{\Omega}^2}{r^2}})}{{d}{t^2}}{+}{{d}{r^2}}}},
\eqno (1)
$$
where $\Omega$ is the pulsar rotation frequency.
Here and below we use so called "geometric units": $c=G=1$.

According to the Einstein principle of equivalence, we can not tell
gravitation from noninertiality. Thus, for the description of the particle
motion in the pulsar magnetosphere the "3+1" formalism can be used. This
formalism is described in [11]. According to the "3+1" formalism, the
equation of the motion for the particle with the mass $m$ and charge $e$ has
the following form [11]:

$$
{{{1}\over {\alpha}}{{{\partial}{\vec p}}\over{{\partial}{t}}}{+}
{({\vec V}{\vec {\nabla}})}{\vec p}{=}{-}{\gamma}
{{{\vec {\nabla}}{\alpha}} \over {\alpha}}{+}
{{e}\over {m}}{({{\vec {E}}+
{[{\vec {V}}{\vec {B}}]}})}}, \eqno (2)
$$
where ${\alpha}$ is the so called "lapse function" and in our case
${\alpha}={\sqrt {1-{{{\Omega}^2}{r^2}}}}$. Here and below we use the
dimensionless momentum $\vec p$ (${\vec p}$ is changed by ${{\vec p}/{m}}$).
We can rewrite the equation (2) for the quantities defined in the
rest inertial frame:

$$
{{{{\partial}{\vec p}}\over {{\partial}{t}}}{+}
{({\vec V}{\vec {\nabla}})}{\vec p}{=}{-}{\gamma}{\alpha}
{\vec {\nabla}}{\alpha}{+}{{e}\over {m}}{({{\vec {E}}+
{[{\vec {V}}{\vec {B}}]}})}}. \eqno (3)
$$

Now let us discuss the motion of the plasma particles in the zeroth
appro\-xi\-mation of the weak turbulence.
In the limits of this approximation the quantities, which are located in the
equation of motion, can be presented as:

$$
{{{\vec E}{=}{\vec {E_0}}{+}{\vec {E_1}}}, \hskip 1cm
{{\vec B}{=}{\vec {B_0}}{+}{\vec {B_1}}}, \hskip 1cm
{{\vec p}{=}{\vec {p_0}}{+}{\vec {p_1}}}}, \eqno (4)
$$
where ${\vec {E_0}}$, ${\vec {B_0}}$ and ${\vec {p_0}}$ are the basic
terms and ${\vec {E_1}}$, ${\vec {B_1}}$ and ${\vec {p_1}}$ are the
perturbations in the first approximation of the expansion over the parameter
of the weak turbulence. The small parameter in the
approximation of weak turbulence for the electron-positron plasma is:

$$
{{{{E_1}^2}\over {{m}{n}{\gamma}}}{\ll}{1}}. \eqno (5)
$$

From the equation of the motion in the zeroth approximation for the radial
acce\-leration one can obtain:

$$
{{{{\partial}{p_{0r}}}\over {{\partial}{t}}}{+}
{({\vec V}{\vec {\nabla}})}{p_{0r}}{=}{-}{\gamma_{0}}
{\Omega}^{2}{r}}.                             \eqno (6)
$$

         Let us introduce new variable $l$, which is connected with
coordinate $r$ by the following relation:

$$
r=l+{{\int}^{t}_{t_0} v(l,\tau)d{\tau}}.    \eqno (7)
$$

         It is clear, that $l$ is a coordinate derived at the moment
$t_0$. Thus, doing the transformation to the Lagrangian variables and using
equation (6), we receive (see also [12]):

$$
{{{d^{2}l}\over{dt^2}}={{{\Omega}^{2}l}\over
{1-{\Omega}^{2}l^2}}{\left[1-{\Omega}^{2}l^2-
2{\left({{dl}\over
{dt}}\right)}^2 \right]}}.                 \eqno (8)
$$

Equation (8) can be solved exactly. Using Jacobian functions, the solution
can be presented in the form [12]:

$$
{{l}{({t})}{=}{{{V_0}_i}\over {\Omega}}{{{Sn}{{{\Omega}{t}}}}\over
{{dn}{{{\Omega}{t}}}}}}, \eqno (9)
$$
where ${Sn}$ and ${dn}$ are the Jacobian elliptical sine and modulus
respectively [13],
${{V_0}_i}$ is the particle initial velocity. From equation (8)
it follows
that, if the radial velocity ${{V_r}>{{1}/{\sqrt {2}}}}$,
the acceleration changes its sign and the
particle is not accelerated, but braked (see also [12]).

In the case ${{V_0}_i}{\rightarrow}{1}$, using the asymptotic expression for
the Jacobian functions one can find [12]:

$$
{{l}{({t})}{=}{{{V_0}_i}\over {\Omega}}{sin}{\Omega}{t}}. \eqno (10)
$$
For the radial velocity we will obtain:

$$
{{{V_0}_r}{=}{{V_0}_i}{cos}{\Omega}{t}}, \eqno (11)
$$
from which it follows that

$$
{{{V_0}^2}{=}{{({{V_0}_r})}^2}{+}{{({{V_0}_{\varphi}})}^2}{=}
{const}} \eqno (12)
$$
(because of the corotation, ${{V_0}_{\varphi}}={\Omega}r$), i.e. no
energy is expending on the particle braking along the field
line, the energy transforms from radial to the transversal one.

\begin{center}
{\large \bf {III. THE EQUATIONS FOR THE PERTURBED QUANTITIES}}
\end{center}

As it was shown above, the relativistic plasma particles are braked
in the pulsar magnetosphere, if their radial velocity ${V_r}>{1/{\sqrt 2}}$.
It is very interesting to discuss the stability of such a plasma with
respect to the radial perturbations. In particular, we discuss the potential
perturbations oriented along the magnetic field lines. The initial stage of
the perturbation deve\-lopment can be described by the equation, which is
easy to obtain from (3) by substituting in it the expansion (4).
For the first order terms one can obtain:

$$
{{{{{\partial}{\vec {p_1}}}\over{{\partial}{t}}}{+}
{({\vec {V_0}}}{\vec {\nabla}})}{\vec {p_1}}{=}
{({\vec {V_0}}{\vec {p_1}})}{{\Omega}^2}{\vec r}{+}
{{e}{\vec {E_1}}}}. \eqno (13)
$$

In order to eliminate the electric field ${\vec {E_1}}$
from the equation (13), let us use the Poisson equation:

$$
{{div}{\vec {E_1}}{=}{4}{\pi}{e}{n_1}}, \eqno (14)
$$
where ${n_1}{=}{n_{{1}{p}}}{-}{n_{{1}{e}}}$ is the perturbed
concentration --- the difference of positron and electron
perturbed concentrations.

Let us complete the equations (13) and (14) with the continuity equation in
the first approximation of the weak turbulence and thus, lock the set of the
equations:

$$
{{{{\partial}{n_1}}\over{{\partial}{t}}}{+}
{div}{n_0}{\vec {V_1}}{+}{div}{n_1}{\vec {V_0}}{=}{0}}. \eqno (13)
$$

Before solving the set of equations (13)-(15), let us introduce the
characteristic spatial parameter of the system and consider the case,
when

$$
{[{{\partial {v_{0r}}} \over {\partial l}}/{v_{0r}}]}^{-1} >> \lambda,
  \eqno (16)
$$
where $\lambda$ is the length of the perturbations. Using the value of
$v_{0r}$ obtained from equation (12), equation (16) could be presented in
the form:

$$
{{{R_{LC}}^{2}-l^2} \over l} >> \lambda,        \eqno (17)
$$
where $R_{LC}=C/{\Omega}$ is the radius of the Light Cylinder.
It follows from equation (17) that, if we investigate the processes in
 the region with the scale
smaller than $R_{LC}$ one, we can consider the medium as homogeneous
one and make the Fourier transformation.
$V_{0}$ - is solved in our description in Lagrangian variables, while the
set of equations (13)-(15) is written in the Eulerian variables. So, let us
express $V_{0r}$ by the variables $r$ and $t$ (see (7)). Doing
the spatial Fourier transformation using approximation (16) for
the perturbed quantities $P_1$, $n_1$
and $E_1$, we can obtain equation, which corresponds to the set of
the eqs. (13)-(15):
$$
{{{\left [{{\partial}\over{{\partial}{t}}}{+}
{{{i}{k_r}{{V_{{0}{r}}}}}}\right ]}^2}{p_{{1}{r}}}{=}
{{{\Omega}{{{V_0}_i}^2}}\over {2}}{\left [{{\partial}\over{{\partial}{t}}}{+}
{{{i}{k_r}{V_{{0}{r}}}}}\right ]}{p_{{1}{r}}}
{sin}{2}{\Omega}{t}{-}
{{{{{\omega}_p}^2}}\over {{{{\gamma}_0}}}}{p_{{1}{r}}}
{{sin}^2}{2}{\Omega}{t}},
\eqno (18)
$$
where ${{\omega}_p}$ is the plasma frequency.

In order to solve the equation (18), it is convenient to introduce an
expression:

$$
{{{k_r}{R}{({t})}{=}{\int}{{k_r}{{V_0}_r}{({t})}}{d}{t}}}. \eqno (19)
$$
Then the equation (18) can be rewritten in the following form:

$$
{{{{\partial}^2}\over {{{\partial}{t}}^2}}{\left (
{exp({{i}{{k}_r}{{R}}})}{{p_1}_r}{({{\vec {r}}{,}{t}})}\right )}{=}
{{{\Omega}{{{V_0}_i}^2}}\over {2}}{{\partial}\over {{\partial}{t}}}{\left (
{exp({{i}{{k}_r}{{R}}})}{{p_1}_r}{({{\vec {r}}{,}{t}})}
{{sin}{2}{\Omega}{t}}\right )}{-}}
$$
$$
{{-}{{{{{\omega}_p}^2}}\over {{{{\gamma}_0}}}}{exp({{i}{{k}_r}{{R}}})}
{{p_1}_r}{({{\vec {r}}{,}{t}})}
{{sin}^2}{\Omega}{t}}. \eqno (20)
$$

Let us expand the function

$$
{exp({{\pm}{i}{{k}_{r}}{{R}}})}{=}
{exp({{\pm}{i}{a}{sin}{\Omega}{t}})}, \eqno (21)
$$
where

$$
{{a}{=}{{{k_r}{{V_0}}}\over {\Omega}}}, \eqno (22)
$$
in the row over the Bessel functions:

$$
{{exp({{\pm}{i}{a}{sin}{\Omega}{t}})}{=}
{\sum_{n=-{\infty}}^{+{\infty}}}{I_n}{({a})}
{exp({{\pm}{i}{n}{\Omega}{t}})}}. \eqno (23)
$$
Here ${I_n}{({a})}$ is a Bessel function. Doing the expansion (23)
in equation
(20)  and using the Fourier transformation in time,
we obtain:

$$
{{{{{{\omega}_p}^2}\over {{{2}{{\gamma}_0}}}}
{{p_1}_r}{({{\vec r}{,}{\omega}})}{=}
{\sum_{{s,n}={-{\infty}}}^{+{\infty}}}{I_s}{({a})}{I_n}{({a})}
{{({{\omega}{-}{n}{\Omega}})}^2}
{{p_1}_r}{({{\vec {r}}{,}{\omega}{+}{({{s}{-}{n}})}{\Omega}})}{-}}}
$$

$$
{-}{{{\Omega}{{{V_0}_i}^2}}\over {4}}
{\sum_{{s,n}={-{\infty}}}^{+{\infty}}}{I_s}{({a})}{I_{n}}{({a})}
{({{\omega}{-}{n}{\Omega}})}
{{p_1}_r}{({{\vec {r}}{,}{\omega}{+}{({{s}{-}{n}{+}{2}})}{\Omega}})}{+}
$$

$$
{+}{{{\Omega}{{{V_0}_i}^2}}\over {4}}
{\sum_{{s,n}={-{\infty}}}^{+{\infty}}}{I_s}{({a})}{I_{n}}{({a})}
{({{\omega}{-}{n}{\Omega}})}
{{p_1}_r}{({{\vec {r}}{,}{\omega}{+}{({{s}{-}{n}{-}{2}})}{\Omega}})}{+}
$$

$$
{{+}{{{{\omega}_p}^2}\over {{4}{{{\gamma}_0}}}}
{{p_1}_r}{({{\vec {r}}{,}{({{\omega}{-}{2}{\Omega}})}})}{+}
{{{{\omega}_p}^2}\over {{4}{{{\gamma}_0}}}}
{{p_1}_r}{({{\vec {r}}{,}{({{\omega}{+}{2}{\Omega}})}})}}, \eqno (24)
$$
where ${\omega}$ is the frequency of the generated perturbations. We will
consider the case of very low frequencies ${\omega}{\ll}{\Omega}$,
which seems to us to be the most interes\-ting. We will assume that
the particle
has no time to change its momentum significantly, which fully corresponds to
the linear app\-roximation of the weak turbulence and describes the initial
stage of the process. In this case all harmonics except the zeroth
${{p_1}_r}{({{\vec r}{,}{\omega}})}$ are small.
So keeping only zeroth harmonics in eq.(18), which corresponds to the
averaging by the fast oscillations and
changing ${n}{\rightarrow}{s}$, we obtain the dispersion relation:

$$
{{{{{\omega}_p}}^2}\over {{{2}{{\gamma}_0}}}}{=}
{\sum_{{s}={-{\infty}}}^{+{\infty}}}{{I_s}^2}{({a})}
{{({{\omega}{-}{s}{\Omega}})}^2}{-}{{{\Omega}{{V_0}^2}}\over {4}}
\biggl [
{({{\omega}{+}{2}{\Omega}})}
{\sum_{{s}={-{\infty}}}^{+{\infty}}}{I_s}{({a})}{I_{s-2}}{({a})}{-}
$$
$$
{-}{\Omega}
{\sum_{{s}={-{\infty}}}^{+{\infty}}}{s}{I_s}{({a})}{I_{s-2}}{({a})}
{{-}{({{\omega}{-}{2}{\Omega}})}
{\sum_{{s}={-{\infty}}}^{+{\infty}}}{I_s}{({a})}{I_{s+2}}{({a})}{+}
{\Omega}
{\sum_{{s}={-{\infty}}}^{+{\infty}}}{s}{I_s}{({a})}{I_{s+2}}{({a})}
\biggl ]}. \eqno (25)
$$

It is easy to find that:

$$
{{\sum_{{s}={-{\infty}}}^{+{\infty}}}{s}{{I_s}^2}{({a})}{=}{0}},
\hskip 1cm
{{\sum_{{s}={-{\infty}}}^{+{\infty}}}{I_s}{({a})}{I_{s{\pm}2}}{({a})}{=}
{0}},
\hskip 1cm
{{\sum_{{s}={-{\infty}}}^{+{\infty}}}{s}{I_s}{({a})}{I_{s{\pm}2}}{({a})}
{=}{0}},
$$

$$
{{\sum_{{s}={-{\infty}}}^{+{\infty}}}{{I_s}^2}{({a})}{=}{1}}
\hskip 0.5cm
and
\hskip 0.5cm
{{\sum_{{s}={-{\infty}}}^{+{\infty}}}{s^2}
{{I_s}^2}{({a})}{=}{{a^2}\over {2}}}.
$$
So from the (25) we can obtain:

$$
{{{\omega}^2}{=}{{{{{\omega}_p}}^2}\over {{{2}{{\gamma}_0}}}}{-}
{{{{k_r}^2}{{{{V_0}_i}}^2}}\over {2}}}. \eqno (26)
$$
We know that ${E_1}{\sim}{exp({{-}{i}{\omega}{t})}}$, so one can
conclude that, when the second term in the right hand side of (26) is
larger than the first term the aperiodic instability is being developed
in the pulsar magnetosphere, i.e. the field ${E_1}$ is increasing
exponentially along the magnetic field lines.

The condition of the aperiodic instability development can be written
in the following form:

$$
{{l_r}<{{{{V_0}_i}{\sqrt {{{\gamma}_0}}}}\over {{\omega}_p}}}, \eqno (27)
$$
where ${l_r}$ is the charge separation scale in the magnetospheric
plasma. For the typical
parameters of the pulsar magnetosphere the charge separation scale at the
light cylinder (the light cylinder is the surface, on which the azimuthal
velocity equals to the speed of light ${V_{\varphi}}={\Omega}r=c$) is
of the order ${{10}^6} sm$. As for the parameter ${({{{V_0}_i}{\sqrt
{{{\gamma}_0}}}})/{{\omega}_p}}$, it differs from the
relativistic generalization of the Debye radius $l_D$ [14] (${l_D}=
{({{V_T}{\sqrt {{\gamma}_T}}})/{{\omega}_p}}$, where ${V_T}$ and
${{\gamma}_T}$ are the particle thermal velocity and Lorentz-factor
respectively).  On the other hand, it is selfevident that the following
condition must be fulfilled:

$$
{{l_r}{\gg}{l_D}}, \eqno (28)
$$
i.e. the charge separation scale $l_r$ must be much larger than the
Debye radius $l_D$. From the conditions (27) and (28) one can conclude
that the aperiodic instability will take place if:

$$
{{{{\gamma}_0}}{\gg}{{{\gamma}_0}_T}}. \eqno (29)
$$

\newpage
\begin{center}
{\large \bf {IV. CONCLUSION}}
\end{center}

At the end let us discuss the possible results of the instability. We
can see that the plasma motion along the magnetic field lines
and at the same time rotation together with them (i.e. corotation)
cause the generation of the aperiodically increasing
electrostatic field under the condition (27).
On the other hand, it is selfevident that the corotation can not take
place on the arbitrary distances from the pulsar surface, because on some
distance the azimuthal velocity will reach the speed of light ${V_{\varphi}}=
{\Omega}r=c$. So, the corotation must be removed.
The instability, which was discussed above, can contribute to the process
of the corotation removing, in particular,
the increasing electric field will cause the
additional braking of the particles of one sort
and the decrea\-sing of bra\-king of
the other sort. This fact will evidently cause
the motion of the electrons and the positrons
with respect to each other, i.e. the increa\-sing current ${\vec j}$ will
appeare. So, according to the Maxwell equ\-ation
${4}{\pi}{\vec j}{=}{rot}{\vec B}$,
the magnetic field will be generated. The current will be
directed along the pulsar magnetic field lines, therefore, the generated
magnetic field will have an azimuthal component ${B_{\varphi}}$.
The particles move along the field
lines, so, the corotation will be removed. The electric field,
i.e. the current ${\vec j}$, will increase untill the corotation law removal.

\begin{center}
{\large \bf {ACKNOWLEDGMENTS}}
\end{center}

G.Z. Machabeli's research was
supported in part by INTAS (Found International Thechnical and Science)
Grant No. 1010 ct 930015.

T.A. Kahniashvili's work was supported in part by ESO C \& EE 
(Scientific and
Thechnical Programs) Grant No. A-05-012.

\begin{flushleft}
\begin{tabular}{ll}
$1.$&P. Goldreich and W.J. Julian, Astrophys. J. {\bf 157}, 869 (1969).\\
$2.$&P.A. Sturrock, Astrophys. J. {\bf 164}, 529 (1971). \\
$3.$&R.N. Henriksen and J.A. Norton, Astrophys. J. {\bf 201}, 719 (1975).\\
$4.$&F.C. Michel, Astrophys. J. {\bf 153}, 717 (1969). \\
$5.$&C.F. Kennel, F.S. Fujimura and I. Okamoto, Geophys. Astrophys. Fluid \\
    &Dynamycs {\bf 26}, 147 (1983).\\
$6.$&L. Mestel, Astrophys. Space Sci. {\bf 24}, 289 (1973).\\
$7.$&J. Arons, Astrophys. J. {\bf 266}, 215 (1983). \\
$8.$&V.G. Endean, Astrophys. J. {\bf 187}, 359 (1974). \\
$9.$&J. Cohen and A. Rosenblum, Astrophys. Space Sci. {\bf 6}, 130 (1972). \\
$10.$&O.V. Chedia, T.A. Kahniashvili, G.Z. Machabeli and I.S. Nanobashvili, \\
     Astrophys. Space Sci. {\bf 239}, 57 (1996)\\
$11.$&"Black Holes: The Membrane Paradigm" eds. K.S. Thorne, R.H. Price,\\
     &D.A. Macdonald (Yale University Press, New Haven and London), 1988.\\
$12.$&G.Z. Machabeli and A.D. Rogava, Phys.Rev. {\bf 50}, 98 (1994).  \\
$13.$&"Handbook of Mathematical Functions", eds. M. Abramowitz and \\
     &I.A. Stegun, 1964. \\
$14.$&D.G. Lominadze, A.B. Mikhailovskii and R.Z. Sagdeev, Zh. Eksp. Teor. \\
     &Fiz. {\bf 77}, 1951 (1979) [Sov. Phys. JETP {\bf 50}, 927 (1979)]. \\
\end{tabular}
\end{flushleft}
\end{document}